\documentclass[12pt,preprint]{aastex}
\usepackage{epsfig}
\newcommand       \Angstrom     {\,{\rm \AA}}

\newcommand       \erg          {\,{\rm erg}}

\newcommand       \simgt        {\gtrsim}

\newcommand       \mum          {\,{\rm \mu m}}

\newcommand       \simali   {\sim\,}
\newcommand       \Alambda  {A_\lambda}
\newcommand       \AV       {A_V}
\newcommand       \Fnu      {F_\nu}
\newcommand       \Fo       {F_{\rm o}}

\newcommand       \magni    {\,{\rm mag}}
\newcommand       \Npara    {N_{\rm para}}
\newcommand       \Ndata    {N_{\rm data}}
\def	\beq	{\begin{equation}}
\def	\eeq	{\end{equation}}
\def	\beqa	{\begin{eqnarray}}
\def	\eeqa	{\end{eqnarray}}
\newcommand{\figwidth}{6.0in}

\shorttitle{Dust in the Hosts of High-Redshift GRBs}

\begin{document}
\title{
%------------- enable for labelling preprint ---------------------------
 \vspace*{-2.0em}
  {\normalsize\rm {\it The Astrophysical Journal Letters}, in press}\\
 \vspace*{1.0em}
       Probing Cosmic Dust of the Early Universe through 
       High-Redshift Gamma-Ray Bursts
%\\{\small DRAFT: \today ~~}
     }
\author{S.~L. Liang and Aigen Li}
\affil{Department of Physics and Astronomy, 
       University of Missouri, Columbia, MO 65211; 
       {\sf shunlinliang@mizzou.edu}, 
       {\sf lia@missouri.edu}
       }
%\slugcomment{Submitted to the Astrophysical Journal}

%\shortauthors{*** \& **}

%\received{}

\begin{abstract}
We explore the extinction properties of the dust 
in the distant universe through the afterglows
of high-redshifted GRBs based on the ``Drude'' model
which, unlike previous studies, does not require 
a prior assumption of template extinction laws.
We select GRB\,070802 at $z\approx 2.45$
(which shows clear evidence for the 2175\,\AA\ extinction bump)
and GRB\,050904 at $z\approx 6.29$, 
the 2nd most distant GRB observed to date.   
We fit their afterglow spectra to determine 
the extinction of their host galaxies.
We find that 
(1) their extinction curves differ 
    substantially from that of the Milky Way, 
    the Small and Large Magellanic Clouds
    (which were widely adopted as template 
    extinction laws in literature);
(2) the 2175\,\AA\ extinction feature appears 
    to be also present in GRB\,050904 at
    $z\approx 6.29$; and
(3) there does not appear to show strong evidence
    for a dependence of dust extinction on redshifts.
The inferred extinction curves are closely reproduced 
in terms of a mixture of amorphous silicate and graphite, 
both of which are expected supernova condensates 
and have been identified in primitive meteorites 
as presolar grains originating from supernovae
(which are considered as the main source of dust at high-$z$).
\end{abstract}


\section{Introduction}
Dust is present in the high-redshift ($z>2$) universe,
as evidenced by the reddening of background quasars,
the depletion of heavy elements in quasar absorption 
systems, and the far infrared (IR) to millimeter (mm)
thermal emission of distant quasars.
Dust plays a crucial role in the formation and evolution 
history of stars and galaxies in the early universe.
The importance of correcting for dust extinction in 
the universe is now widely recognized.
In order to reveal the structure and evolution 
of the early universe, 
to use Type Ia supernovae (SNe) as standard candles,
and to infer the cosmological star formation rate, 
it is essential to correct for the effects of dust extinction.

Gamma-ray bursts (GRBs), owing to their intense luminosity
(emitting up to $\simali$$10^{53}\erg$), allow their detection 
up to very high redshifts at $z\simgt 10$ (Lamb \& Reichart 2000).
Particularly, the association of long-duration bursts with massive 
stars (and therefore with dusty regions of high-mass star formation) 
and the featureless, power law-like spectral shapes of 
their afterglows, make GRBs an excellent probe of the dust
at high-redshifts.

In this {\it Letter} we explore the dust extinction of 
the host galaxies of 
GRB\,070802 at $z\approx 2.45$ and
GRB\,050904 at $z\approx 6.29$.
We aim at a quantitative examination of the nature
of the dust in the early universe and attempt to address
one of the hotly-debated questions in high-$z$ astrophysics: 
do the dust properties evolve as a function of redshift
(particularly at $z>5$ where the dust source may be different)?

\section{Dust Extinction Model}
We characterize the dust extinction properties of
GRB hosts with the extinction quantity
(e.g. $A_{V_r}$, the rest-frame visual extinction)
and the wavelength-dependence of the extinction 
(i.e. $A_\lambda/A_V$ or $A_\nu/A_V$ if expressed 
in frequency $\nu$, often known as 
the ``extinction curve'' or ``extinction law''). 
We derive $A_{V_r}$ and $A_\lambda/A_V$ (or $A_\nu/A_V$)
by fitting the ultraviolet (UV), optical, and near-IR 
afterglow photometry with a dust-reddened power-law model 
through
\begin{equation}
\label{eq:Fnu}
\Fnu = \Fo\,\left(\nu/{\rm Hz}\right)^{-\beta} 
\exp\left[-\frac{A_{V_r}}{1.086} \frac{A_{(1+z)\nu}}{A_{V_r}}\right]~~,
\end{equation}
where $F_\nu$ is the afterglow photometry 
(with the Galactic foreground extinction corrected),
$\beta$ is the intrinsic power-law slope 
of the afterglow, 
$\Fo$ is a normalization constant, 
$A_{(1+z)\nu}$ is the rest-frame extinction,
and $z$ is the GRB redshift.

Unlike previous studies which often assume a {\it template} 
extinction law for $A_\lambda/A_V$ (i.e., the extinction
curves of GRB hosts are assumed to resemble that of 
the Milky Way [MW], the Small Magellanic Cloud [SMC],
the Large Magellanic Cloud [LMC], 
the ``Calzetti'' attenuation law of 
starbust galaxies [Calzetti et al.\ 1994],
or the relatively flat ``Maiolino'' curve 
of AGNs [Maiolino et al.\ 2001]),
we take the ``Drude'' model proposed in Li et al.\ (2008a).
This approach approximates the wavelength-dependence 
of the extinction by a simple formula consisting of 
four dimensionless parameters 
($c_1$, $c_2$, $c_3$, and $c_4$)
%%%
\beqa
\nonumber
\label{eq:A2AV}
\Alambda/\AV & = & \frac{c_1}{\left(\lambda/0.08\right)^{c_2}
+ \left(0.08/\lambda\right)^{c_2} + c_3} \\
\nonumber
& + & \frac{233\left[1 - c_1/\left(6.88^{c_2}+0.145^{c_2}+c_3\right)
- c_4/4.60\right]}{\left(\lambda/0.046\right)^2
+ \left(0.046/\lambda\right)^2 + 90} \\
& + & \frac{c_4}{\left(\lambda/0.2175\right)^2
+ \left(0.2175/\lambda\right)^2 -1.95} ~~~,
\eeqa
%%%
where $\lambda$ is in $\mu$m,
the first term in the right-hand side
represents the far-UV extinction rise, 
the second term and the third term respectively 
account for the near-IR/visible
extinction and the 2175\,\AA\ extinction bump.

Compared to models based on template extinction curves,
the ``Drude'' model is preferred because 
(1) it eliminates the need for a prior assumption of 
    template laws -- after all, there is no reason to 
    assume that the ``true'' extinction curves of GRB
    hosts should resemble any of those templates, and 
(2) the analytical formula (eq.\ref{eq:A2AV}) on which
    the ``Drude'' model is based restores 
    the widely-adopted MW, SMC, LMC, ``Calzetti'', 
    and ``Maiolino'' templates -- if the ``true'' extinction 
    curve of a GRB host happens to resemble a certain template 
    law, the ``Drude'' approach will allow us to restore it 
    (see Li et al.\ 2008a).

\section{Results}
We apply the ``Drude'' model to 
GRB\,070802 at $z\approx 2.45$ and
GRB\,050904 at $z\approx 6.29$.
They are selected for the following reasons:
(i) they span a wide range of redshifts,
    from the moderately high redshift of 
    $z\approx 2.45$ (GRB\,070802) 
    to the 2nd highest redshift observed to date
    of $z\approx 6.29$ (GRB\,050904);
(ii) the afterglow photometry of GRB\,070802 provides 
    the most definite evidence for the presence of 
    the 2175\,\AA\ extinction feature in a GRB 
    host galaxy (Kr\"uhler et al.\ 2008;
    El\'iasd\'ottir et al.\ 2008); and
(iii) the peculiar UKIRT $z$ band 
     ($\lambda_{\rm rest}\approx 1275\Angstrom$)
     flux suppression of the GRB\,050904 afterglow
     at 0.5 days and 1 day after the burst 
     (Haislip et al.\ 2006; Stratta et al.\ 2007) 
     was interpreted as evidence for an evolution
     of the dust properties at $z>6$
     (Stratta et al.\ 2007).
 
Using eqs.(\ref{eq:Fnu},\ref{eq:A2AV}) 
and the Levenberg-Marquardt minimization 
algorithm, we fit the broadband 
spectral energy distributions (SEDs) of the afterglows 
of these GRBs\footnote{%
  For GRB\,050904 we will consider three different 
  epochs after the burst.
  }
with $\beta$, $A_V$, $c_1$, $c_2$, $c_3$ and $c_4$ 
allowed to vary as free parameters.\footnote{%
  $\Fo$ is not really a free parameter; 
  for a given set of ($\beta$, $A_V$, $c_1$, $c_2$, $c_3$, $c_4$), 
  $\Fo$ is uniquely determined by the overall flux level.
  } 
Therefore, in the SED modeling we have 
six free parameters.\footnote{%
  Admittedly, the models based on template extinction laws
  have fewer parameters: with the shape of the extinction 
  curve fixed, they only need to determine $\beta$ and $A_V$.
  The ``Drude'' approach needs four more parameters
  (i.e. $c_1$, $c_2$, $c_3$ and $c_4$) to describe 
  the wavelength-dependence of the extinction.
  This is the nature of the ``Drude'' approach; 
  because of this the ``Drude'' approach is more flexible
  in revealing the ``true'' extinction curve.
  }
It is unfortunate that the number of model parameters 
($\Npara = 6$) exceeds the number of photometry data 
points $\Ndata$ for GRB\,050904 
($\Ndata = 4$ for all three epochs; Haislip et al.\ 2006;
Tagliaferri et al.\ 2005).
With $\Ndata = 7$, GRB\,070802 has a better wavelength 
coverage.\footnote{%
  For GRB\,070802, we adopt the optical and near-IR 
  photometry of Kr\"uhler et al.\ (2008) obtained by 
  the 7-channel Gamma-Ray Burst Optical and Near-IR Detector 
  ({\it GROND}) mounted on the 2.2\,m ESO/MPI Telescope.
  The ESO VLT spectroscopy of GRB\,070802 is in close
  agreement with the GROND photometry
  (see Fig.\,5 of El\'iasd\'ottir et al.\ 2008).
  }
We therefore use $\chi^2/\Ndata$ as a quality measure 
of the fit.

In Figure 1 we plot the ``Drude'' model fit to
the afterglow SED of GRB\,070802
as well as the derived extinction curve.
The results for GRB\,050904 at three different
epochs after the burst are shown in Figure 2.
We see in these figures that 
(1) the ``Drude'' model provides excellent fits to 
    the observed SEDs;
(2) the derived extinction curves differ 
    substantially from the widely-adopted 
    template extinction laws;
(3) the 2175\,\AA\ extinction feature appears 
    to be also present in the afterglow spectra 
    of GRB\,050904, the 2nd most distant GRB 
    observed to date, at epochs of 0.5 days 
    and 1 day after the burst;
(4) at an epoch of 3 days after the burst,
    the 2175\,\AA\ feature appears to be absent
    in GRB\,050904, suggesting that its carrier
    may have been destroyed by the burst;\footnote{%
       Indeed, one sees in Figure 2 a gradual flattening
       of the far-UV extinction rise from 0.5\,days 
       to 1\,day and 3\,days after burst, as expected
       from a preferential destruction of small grains
       responsible for the far-UV extinction
       by the burst (see Perna et al.\ 2003),
       that is reflected in Table 2
       with a gradual increase (decrease) 
       of the cutoff sizes 
      (the power-law size distribution indices).
      }
and (5) there does not appear to show strong evidence
    for a dependence of dust extinction on redshifts
    (although the extinction curve does vary from 
    one burst to another), as supported by 
    a systematic study of $>$\,20 GRBs at $z>2$:
    the overall wavelength dependence of extinction,
    the steepness of the far-UV extinction rise,
    and the presence and strength of the 2175\,\AA\ 
    extinction bump, do not appear to show any dependence 
    on redshifts (S.L. Liang \& A. Li 2008, in preparation).
The model parameters are tabulated in Table 1.

\section{Discussion}
In deriving the extinction of GRB hosts,
a major problem with the models based on template
extinction laws is that the wavelength-dependence
of the extinction is fixed. 
For a featureless, power-law-like afterglow SED, 
this often leads to a preference of a SMC-type 
extinction and a small amount of $A_V$ 
(usually $<0.2\magni$): obscured by a SMC-type 
extinction (which is roughly a power-law 
$A_\lambda\propto \lambda^{-1.2}$), 
an intrinsic power-law-like afterglow SED
remains featureless and becomes a steeper power-law. 
However, if the dust is ``gray'' (i.e. the extinction 
$A_\lambda$ only weakly varies with $\lambda$), 
the resulting dust-obscured afterglow SED will still
be a featureless power-law, with the intrinsic
power-law exponent unchanged. 
The possible presence of gray extinction has been 
suggested by a number of authors
(e.g. see Savaglio et al.\ 2003, 
Savaglio \& Fall 2004, Stratta et al.\ 2004, 2005, 
Chen et al.\ 2006, Li et al.\ 2008b, Perley et al.\ 2008).
The ``Drude'' approach allows us to break the degeneracy
between ``gray'' extinction and SMC-type extinction.

Attempts have also been made to fit the afterglow SEDs 
with the MW, SMC and LMC template extinction laws.
As shown in Figures 1,2, no acceptable fits 
are obtained, except that the MW model for GRB\,070802 
and the SMC model for GRB\,050904 at an epoch of 
3 days after the burst fit the observed SEDs reasonably well.
But even for these two cases, the ``Drude'' 
approach fits better as can be seen in Figs.\,1,2 
and indicated by $\chi^2/N_{\rm data}$ (see Table 1).

El\'asd\'ottir et al.\ (2008) tried to fit the VLT/FORS2
spectroscopy and the GROND photometry with 
the Fitzpatrick \& Massa (1990; hereafter FM)
parametrization as well as the MW-, LMC-, and SMC-type 
extinction. They found that satisfactory fits 
could be achieved only if one assumes a cooling break
in the intrinsic spectrum, with the FM parametrization
providing the best fit. However, one should caution that
the FM parametrization is only valid for $\lambda< 2700$\,\AA,
while the GROND photometry of GRB\,070802
extends from $\simali$1400\,\AA\ to $\simali$6400\,\AA\
(in the GRB rest-frame).

The afterglow SEDs of the bursts discussed here
all show a flux suppression at $\lambda \sim 4-6\mum^{-1}$
and deviate appreciably from a power-law
(except for GRB\,050904 at 3 days after the burst).
As shown in Figures 1,2, the flux suppression is 
closely accounted for in terms of dust with 
a 2175\,\AA\ bump in its extinction. 
For GRB\,070802, the derived 2175\,\AA\ bump is 
comparable or even slightly stronger 
than that of the MW: for MW $c_4\approx 0.05$
while $c_4\approx 0.06$ for GRB\,070802.
To validate the suggested detection of the 2175\,\AA\ 
extinction feature, we have also tried to fit 
the afterglow SEDs with the ``Drude'' approach 
but setting $c_4=0$ (i.e. no 2175\,\AA\ extinction bump).
It is found that the fits (with $c_4=0$) are much worse,
as reflected from the substantially increased 
$\chi^2/N_{\rm data}$ (see Table 1). 

The 2175\,\AA\ bump, first detected by
Stecher (1965), is the strongest spectroscopic 
interstellar extinction feature.
This feature is seen in extinction
curves along lines of sight in the MW and LMC.\footnote{%
  Most SMC extinction curves have no detectable 
  2175\,\AA\ bump (Pr\'evot et al.\ 1984).
  But there exist regional variations 
  in the SMC extinction curve.
  The SMC sight lines which show no 2175\,\AA\ bump 
  all pass through the SMC Bar regions of active
  star formation (Pr\'evot et al.\ 1984; 
  Gordon \& Clayton 1998). 
  The 2175\,\AA\ bump is seen at least in one line of sight,  
  Sk 143 (AvZ 456), which passes through the SMC wing,
  a region with much weaker star formation 
  (Gordon \& Clayton 1998).
  }
But it is rarely seen in the afterglow spectra of GRBs.
So far, its possible detection is only reported in
four bursts: GRB\,970508 (Stratta et al.\ 2004),
GRB\,991216 (Kann et al.\ 2006, Vreeswijk et al.\ 2006), 
GRB\,050802 (Schady et al.\ 2007),
and GRB\,070802 (Fynbo et al.\ 2007, Kr\"uhler et al.\ 2008),
with the latter showing the clearest presence of
the 2175\,\AA\ extinction feature in its afterglow spectrum.
In addition, Ellison et al.\ (2006) reported the detection
of this feature in an intervening absorber at $z\approx 1.11$
toward GRB\,060418. But the host galaxy of GRB\,060418 
at $z\approx 1.49$ seems to have a SMC-type extinction law.

The possible detection of the 2175\,\AA\ extinction feature 
has been reported for a number of low, intermediate,
and moderately high redshift systems through 
(1) the composite absorption spectrum of 
    intervening MgII absorption systems 
    (Malhotra 1997: $0.2<z<2.2$) 
    or radio galaxies 
    (Vernet et al.\ 2001: $z\sim 2.5$);\footnote{%
       But York et al.\ (2006) found no evidence for 
       the 2175\,\AA\ bump in the composite 
       absorption spectra of 809 intervening QSO 
       MgII absorbers at $1<z<1.9$.
       }
(2) the individual absorption spectra of 
    intervening MgII absorbers
    (Wang et al.\ 2004: $1.4<z<1.5$;
     Srianand et al.\ 2008: $z\sim 1.3$);\footnote{%
       The 2175\,\AA\ extinction feature, 
       the 9.7$\mum$ silicate absorption feature,
       and the diffuse interstellar bands are seen
       in the damped Ly$\alpha$ absorber at $z\approx0.524$ 
       toward the BL lac object AO\,0235+164 
       (Junkkarinen et al.\ 2004, Kulkarni et al.\ 2007).
       }
(3) the UV SEDs of massive, UV-luminous star-forming 
    galaxies (Noll \& Pierini 2005: $2<z<2.5$;
    Noll et al.\ 2007: $1<z<2.5$); and   
(4) the extinction curves of gravitational lensing
    galaxies (Toft et al.\ 2000: $z\approx0.44$;
    Motta et al.\ 2002: $z\approx0.83$; 
    Wucknitz et al.\ 2003: $z\approx0.93$; 
    Mu\~noz et al.\ 2004: $z\approx0.68$). 
However, Vijh et al.\ (2003) found that the dust 
in 906 Lyman break galaxies at $2<z<4$ does not
exhibit the 2175\,\AA\ extinction feature.
This is probably related to the survival and destruction
of the carriers of the 2175\,\AA\ bump in different
physical conditions. 

Although the precise nature of the carrier of 
the 2175\,\AA\ extinction feature remains unknown, 
it is generally accepted that it arises from small 
graphitic dust or a cosmic mixture of polycyclic 
aromatic hydrocarbon (PAH) molecules (Li \& Draine 2001). 
In view of the detection of presolar graphite dust
with a SN origin in primitive meteorites,
it is not unreasonable to expect a 2175\,\AA\ extinction
bump for high-$z$ objects since the dust at $z>5$ is
thought to originate from Type II SNe.
On the other hand, PAHs have been detected in ultraluminous
IR galaxies and submm galaxies at $z>2$ through their
vibrational bands at 6.2, 7.7, 8.6 and 11.3$\mum$
(see Lutz et al.\ 2005, Yan et al.\ 2005).
PAHs were also seen in the Cloverleaf lensed QSO 
at $z\approx 2.56$ (Lutz et al.\ 2007).   
If PAHs are indeed responsible for the 2175\,\AA\
extinction, it would not be surprising to see this
feature in high-$z$ galaxies.

Finally, we fit the inferred extinction curves 
using a mixture of spherical amorphous silicate 
and graphite dust each with an exponential-cutoff 
power-law size distribution (e.g. see Kim et al.\ 1994)
\begin{eqnarray}
A_{\lambda}/A_{V}=
&  & A_{\rm sil}\int_{a_{\rm min}}^{a_{\rm max}}\,
     C_{\rm ext}^{\rm sil}(a,\lambda)\,
     a^{-\alpha_{\rm sil}}\,
     \exp\left(-a/a_{c,\rm sil}\right)\,da\nonumber \\
&  & + A_{\rm gra}\int_{a_{\rm min}}^{a_{\rm max}}\,
       C_{\rm ext}^{\rm gra}(a,\lambda)\,
       a^{-\alpha_{\rm gra}}\,\exp\left(-a/a_{c,\rm gra}\right)\,da,
\end{eqnarray}
where the lower (upper) cutoff size
$a_{\rm min}$ ($a_{\rm max}$) is taken to be
50\,\AA\ (1$\mum$) for both silicate and graphite dust;
the power-law indices $\alpha_{\rm sil}$, $\alpha_{\rm gra}$ 
and the exponential-cutoff sizes $a_{\rm c,sil}$
and $a_{\rm c,gra}$ are treated as free parameters;
$A_{\rm sil}$ and $A_{\rm gra}$ are related to 
the abundance of each species;
and $C_{\rm ext}^{\rm sil}$ 
($C_{\rm ext}^{\rm gra}$) is the extinction cross section 
of silicate (graphite) dust.
As shown in Figure 1c and Figure 3, 
the silicate-graphite model closely
reproduces the inferred extinction curves for both GRBs, 
including the 2175\,\AA\ extinction bump
(see Table 2 for the size parameters).
The major mismatch occurs at $\lambda\sim 7\mum^{-1}$
which is probably due to the sudden rise of the silicate
electronic absorption (see Kim \& Martin 1995).
We note that both silicate and graphite 
are expected SN condensates
(Todini \& Ferrara 2001, Nozawa et al.\ 2003).
They have been identified as presolar grains
in primitive meteorites originating from supernovae
which are considered as the main source of dust 
at $z>5$ (see Dwek et al.\ 2007). 

By fitting the afterglow SEDs of GRB\,050904 
($z\approx 6.29$) with the extinction curve inferred 
for the distant BAL QSO at $z\approx 6.2$
(which displays a plateau at $\lambda^{-1}\sim 3.3 - 5.9\mum^{-1}$,
Maiolino et al.\ 2004), Stratta et al.\ (2007) 
argued that the dust properties may evolve beyond $z>5$. 
This seems to be supported by that the dust at $z>5$ 
is probably produced by Type II SNe
while in the local universe AGB stars are 
a major source of dust.
However, this study together with a preliminary analysis
of $>$\,20 GRBs at $z>2$ based on the ``Drude'' approach
does not indicate any dependence of the dust extinction 
on redshift. A more thorough and systematic study
of the dust extinction and IR emission properties
of high-$z$ GRBs is in progress and will be used to
further explore whether the dust properties vary as 
a function of redshift.

\acknowledgments
We thank V.P. Kulkarni, S. Savaglio, G. Stratta, D. Watson 
and D.M. Wei for very helpful comments and suggestions.
We are supported in part by a NASA/Swift Theory Program, 
a NASA/Chandra Theory Program, and the NSFC Outstanding 
Oversea Young Scholarship.

% -------------------------------------------------------------------------

%*********** TABLE 1 *****************
%\clearpage
\begin{table}
\caption[]{\footnotesize
           Parameters for fitting the afterglow SEDs
           with the ``Drude'' model
           and the MW, LMC and SMC template extinction laws.
           \label{tab:grbmod}}
\begin{center}
{\scriptsize
\begin{tabular}{lccccccccc}
\tableline\tableline
Extinction & $c_1$ & $c_2$ & $c_3$ & $c_4$
           & $\AV$ & $\beta$ & $\Fo$  
           & $\chi^2/N_{\rm data}$  \\
Type       &  &  &   &   & (mag) &         & ($\mu$Jy) &    \\
\cline{1-10}
& & \multicolumn{6}{c}{GRB\,070802 ($z\approx 2.54$)} & &\\
\cline{1-10}
{\bf Drude}         &{\bf 0.08}  &{\bf 0.32} &{\bf -1.99}
                    &{\bf 0.06}  &{\bf 0.81} &{\bf 0.98}
                    &{\bf 2.38E17} &{\bf 0.23}\\
Drude         &0.10  &0.34 &-1.98 &0.00 &0.83 &0.97 &1.70E17 &1.86 \\
MW    & ...   & ...   & ...   & ... &0.81 &1.39 &9.39E22 &0.84 \\
SMC   & ...   & ...   & ...   & ... &0.91 &1.09 &3.80E18 &3.43  \\
LMC   & ...   & ...   & ...   & ... &1.57 &0.002 &628.3 &0.66\\
\cline{1-10}
& & \multicolumn{6}{c}{GRB\,050904 ($z\approx 6.29$; 
                       0.5\,days after burst)} & &\\
\cline{1-10}
{\bf Drude}         &{\bf 0.91}   &{\bf 1.62}  &{\bf -2.34}
                    &{\bf 0.02}   &{\bf 0.38}  &{\bf 0.25}
                    &{\bf 7.37E8} &{\bf 0.01} \\
Drude         &0.86 &1.73 &-2.30 &0.00 &0.42 &0.27  &1.43E9 &1.20\\
MW    & ...   & ...   & ...   & ... &0.01     &1.42    &1.12E26   &5.51  \\
SMC   & ...   & ...   & ...   & ... &0.41     &0.001   &1.13E5   &2.37  \\
LMC   & ...   & ...   & ...   & ... &0.46     &0.35    &2.23E10   &3.94\\
\cline{1-10}
& &  \multicolumn{6}{c}{GRB\,050904 ($z\approx 6.29$; 
                        1\,day after burst)}  & & \\
\cline{1-10}
{\bf Drude}         &{\bf 1.31}   &{\bf 1.07}   &{\bf -1.99}
                    &{\bf 0.03}   &{\bf 0.39}   &{\bf 0.24}
                    &{\bf 4.80E8} &{\bf 0.01}\\
Drude         &1.54  &1.13 &-1.99 &0.00 &0.46 &0.26  &9.41E8 &1.12 \\
MW    & ...   & ...   & ...   & ... &0.16   &1.73    &6.11E30  &5.58 \\
SMC   & ...   & ...   & ...   & ... &0.53   &0.001   &1.24E5  &1.61\\
LMC   & ...   & ...   & ...   & ... &0.84   &0.03    &6.01E5  &2.07\\
\cline{1-10}
& & \multicolumn{6}{c}{GRB\,050904 ($z\approx 6.29$; 
                       3\,days after burst)} & & \\
\cline{1-10}
{\bf Drude}         &{\bf 1.58}   &{\bf 1.18}   &{\bf -1.72}
                    &{\bf 0.00}   &{\bf 0.41}   &{\bf 0.26}
                    &{\bf 1.95E8} &{\bf 0.04}\\
%Drude($c_4=0$)&1.66 &1.35 &-1.83 &0.00 &0.43 &0.28 &3.73E9  &0.14 \\
%
MW    & ...   & ...   & ...   & ... &0.001   &1.34  &1.19E24    &0.61 \\
SMC   & ...   & ...   & ...   & ... &0.33    &0.16  &4.10E6    &0.06 \\
LMC   & ...   & ...   & ...   & ... &0.24    &0.79  &1.01E16    &0.39 \\
\tableline
\end{tabular}
}
\end{center}
\end{table}

%*********** TABLE 2 *****************
%\clearpage
%\begin{deluxetable}{l r | c c c | c c c}
\begin{deluxetable}{l r c c c c c c}
%\rotate{}
\tablecolumns{10}
\tabletypesize{\scriptsize}
\tablewidth{0pc}
\tablecaption{Dust size distributions for
              the extinction curves derived 
              from the ``Drude'' model and
              a mixture of silicate and graphite grains}
\tablehead{
\colhead{GRB} &
\colhead{$z$} &
\colhead{$A_{\rm sil}$} &
\colhead{$\alpha_{\rm sil}$} &
\colhead{$a_{\rm c,sil}$\,($\mu$m)} &
\colhead{$A_{\rm gra}$} &
\colhead{$\alpha_{\rm gra}$} &
\colhead{$a_{\rm c,gra}$\,($\mu$m)}
}
\startdata
070802
&2.45 &0.30   &2.84     &0.039     &0.70     &3.03      &0.11 \\
050904\,(0.5\,days)
&6.29 &0.59   &3.08     &0.014     &0.41     &3.10      &0.33 \\
050904\,(1\,day)
&6.29 &0.63   &3.05     &0.021     &0.37     &3.08      &0.52 \\
050904\,(3\,days)
&6.29 &0.68   &3.00     &0.045     &0.32     &2.88      &0.76 \\
\enddata
%\tablenotetext{}{}
%\label{tabALL}
\end{deluxetable}

%%%%%%%%%%%%%%%%%%%%%%%%%%%%%%%%%%%%%%%%%%%%%%%%%%%%%%%%%%%%%%%%%

\begin{figure}[ht]
\begin{center}
\includegraphics[width=\figwidth,angle=0]{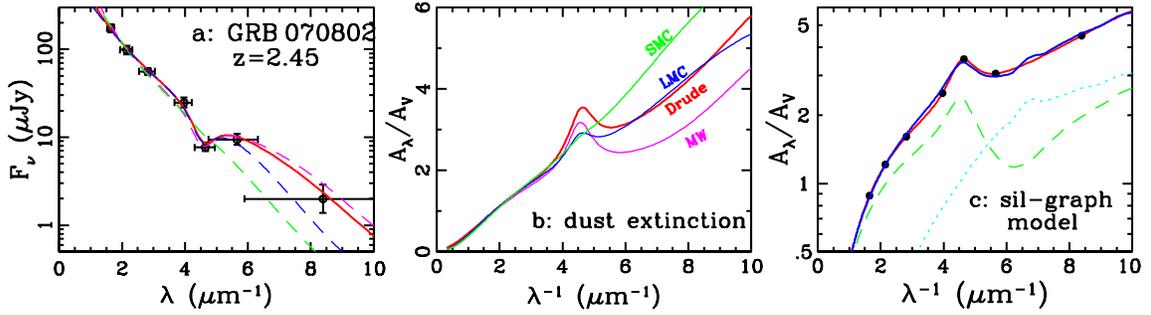}
\end{center}\vspace*{-1em}
\caption{
        \label{fig:extcurv1}
        Left panel (a): Fitting the SED of the afterglow
        of GRB\,070802 ($z\approx 2.45$)
        with the ``Drude'' approach (red)
        and the MW (magenta), LMC (blue) and SMC (green)
        templates for the GRB host extinction curve.
        Middle panel (b): Comparison of the MW (magenta),
        LMC (blue), and SMC (green) extinction laws with
        that derived from the Drude approach (red).
        Right panel (c): Fitting the derived extinction curve 
        (red solid line and black filled circles) 
        with a mixture of amorphous silicate (cyan dotted line) 
        and graphite dust (green dashed line).
        The blue solid line plots the resulting model extinction curve.
        }
\end{figure}

\begin{figure}[ht]
\begin{center}
\includegraphics[width=\figwidth,angle=0]{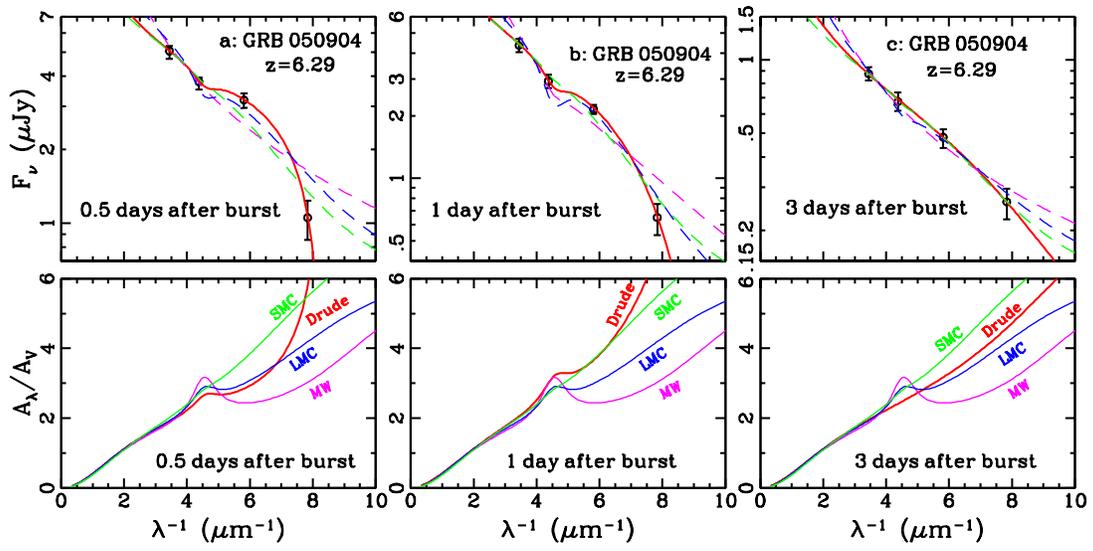}
\end{center}\vspace*{-1em}
\caption{
        \label{fig:extcurv2}
        Same as Figure 1a,b but for GRB\,050904 
        (Haislip et al.\ 2006; Tagliaferri et al.\ 2005)
        at three different epochs after burst.
        }
\end{figure}

\begin{figure}[ht]
\begin{center}
\includegraphics[width=\figwidth,angle=0]{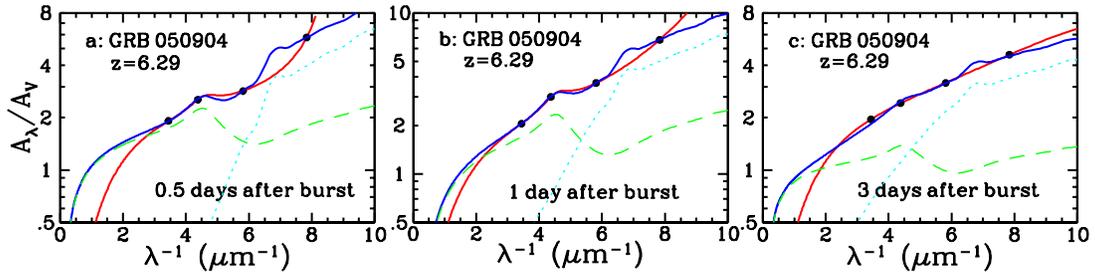}
\end{center}\vspace*{-1em}
\caption{
        \label{fig:sizedistrib}
        same as Figure 1c but for GRB\,050904 
        at three different epochs after burst. 
        }
\end{figure}

\end{document}